\documentclass[a4paper,12pt]{article}
\usepackage{amsfonts,amssymb,amsmath}
\usepackage{bm,euscript,mathrsfs}
\usepackage[utf8]{inputenc}
\usepackage[T1]{fontenc}
\usepackage{color,soul}
\usepackage{textcomp}
\usepackage[tight,sf]{subfigure}
\usepackage{hyperref}
\usepackage{graphicx}
\usepackage{color}
\usepackage{psfrag}

\title{\Huge{\bf Confining spheres within hyperspheres}}
\date{}
\author{\Large
Jemal Guven${}^1$\footnote{\href{mailto:jemal@nucleares.unam.mx}{
jemal@nucleares.unam.mx}}\,, Jos\'e Antonio Santiago${}^2$\footnote{jsantiago@correo.cua.uam.mx} and
Pablo V\'azquez-Montejo${}^2$\,\footnote{\href{mailto:pvazquez@correo.cua.uam.mx}{
pvazquez@correo.cua.uam.mx}}}

\begin{document}
 \maketitle
\begin{center}
{\it
$1$ Instituto de Ciencias Nucleares,
 Universidad Nacional Aut\'onoma de M\'exico\\
 Apdo. Postal 70-543, 04510 M\'exico, D.F., MEXICO
\\
$2$ Departamento de Matem\'aticas Aplicadas y Sistemas,\\
Universidad Aut\'onoma Metropolitana-Cuajimalpa, C.P. 01120, M\'exico D.F., MEXICO}
\end{center}

\begin{abstract}
The bending energy of any freely deformable closed surface is
quadratic in its curvature. In the absence of constraints, it
will be minimized when the surface adopts the form of a
round sphere. If  the surface is confined within a hypersurface of
smaller size, however, this spherical state becomes inaccessible.
A framework is introduced to describe the equilibrium states of
the confined surface. It is applied to a two-dimensional surface
confined within a three-dimensional hypersphere of smaller radius.
If the  excess surface area is small,  the equilibrium
states are represented by harmonic deformations of a
two-sphere: the ground state is described by a quadrupole;
all higher multipoles are shown to be unstable.
\end{abstract}

PACS numbers: 45.10.Db, 02.40.-k, 87.16.D-

\section{\bf Introduction}

\begin{figure}[htb]
\begin{center}
 \includegraphics[scale=0.12]{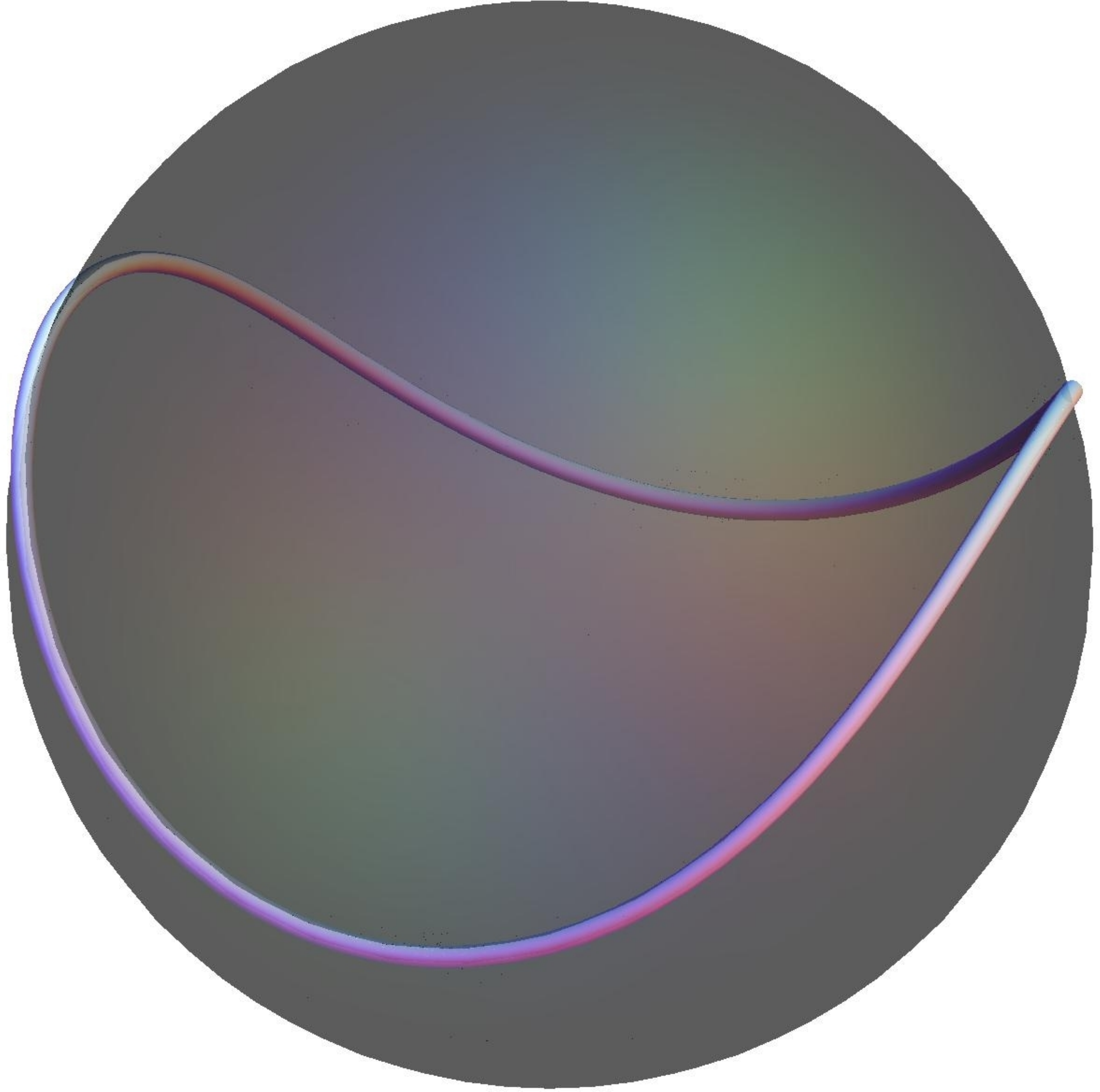}
\end{center}
\caption{\small{Equilibrium state of an elastic loop with two-fold symmetry}}
\label{fig1}
\end{figure}

Twenty five years ago, Nickerson and Manning examined how an elastic curve minimizes its energy when
its freedom to bend in space is constrained by a two-dimensional obstacle \cite{ Mann, MannNick}.
Their motivation was to understand the wrapping of DNA, a semi-flexible polymer, around histone
cylinders but the question is also  relevant to an understanding of the confinement
of DNA within viral capsids (see, for example, \cite{Spakowitz}).
\vskip1pc \noindent
Whereas a space curve minimizes the Euclidean distance between two points by negotiating obstacles
along surface geodesics,  an elastic curve generally will not. This is because the  relevant energy
is not arc-length but the bending energy of the curve in three-dimensional space,
quadratic in the spatial Frenet curvature $k$ \cite{Kamien}. This energy is always greater than its
surface counterpart, quadratic in the geodesic curvature $\kappa_g$, characterising  how the curve
bends within the surface. By projecting the curvature vector of the space curve along the
surface and normal to it, its bending energy can be decomposed as a sum of two contributions:
$k^2 = \kappa_g^2 + \kappa_n^2$  \cite{DoCarmo}; whereas the intrinsic $\kappa_g$ is insensitive to
the three-dimensional environment,  the normal curvature $\kappa_n$ will register how the surface
itself bends in space.  When the surface is a sphere, $\kappa_n$ is a constant and thus the spatial
bending energy will be minimized by surface geodesics. The snag, however, is that the boundary
conditions, and in particular the closure of the curve, tend to be incompatible with geodesic
behavior.
\vskip1pc \noindent
In \cite{GuvVaz} the equilibrium states of a closed elastic loop confined within a sphere were
examined. It was found that, if the loop is short,  there is a unique ground state;
it is completely attached and it exhibits a two-fold dihedral symmetry, shown in Fig. \ref{fig1}.
This state is stable; all of its excited counterparts are unstable. The bending energy was also
found to depend--in a surprisingly sensitive way--on the length of the loop. Contradicting naive
expectations, neither the energy nor the forces transmitted to the sphere increase monotonically
with loop length. Local maxima are associated with the incommensurability of loop
length with the equatorial circumference.
\vskip0pc \noindent
In this paper, higher dimensional analogues of a confined elastic loop will be examined. In
particular, the focus will be on confined two-dimensional surfaces. The surface energy is now the
two-dimensional bending energy, quadratic in its curvature (see, for example, \cite{Willmore, Palmer,
LandauLifshitz, Polyakov, canhamHelfrich, Seifert}). In the absence of constraints, it is simple to
demonstrate that the closed surface minimizing this energy will be a round sphere.
\vskip1pc \noindent
The obvious higher-dimensional analogue of a confined loop is
a two-dimensional closed surface confined within a sphere of smaller area.
This is a problem of considerable interest from a biological
perspective--where the morphology of cellular compartments may involves the confinement of one
membrane within another--and it has been explored recently by Kahraman et al. \cite{Muller EPL,
Muller New Journal}.  In distinction to the loop,  which attaches everywhere to the sphere,  the
confined surface must detach somewhere, folding  into the cavity.  The interesting behavior
occurs within these interior folds and along the boundaries of the regions of contact, not within
the regions of contact themselves.
\vskip1pc \noindent
A more appropriate analogue of the confined loop is provided by locating the surface in a higher
dimensional environment.  The simplest possibility is to embed the two-dimensional surface in a
four-dimensional Euclidean geometry, confining it within a three-dimensional hypersurface,  and, in
particular, a hypersphere of smaller radius. While this may appear a somewhat recondite exercise from
the practical--application-driven--point of view of condensed matter, geometrically it is the natural
generalization of the confined loop. Mathematically, it also can be viewed as a natural
generalization of the Willmore problem involving competing length scales\cite{Willmore, Palmer}.
Indeed, when the constraint on the area is
relaxed, it reduces to the Willmore problem as formulated on a three-sphere, $\mathbb{S}^3$.
\vskip1pc \noindent
To address this problem, the constraint is imposed by introducing a local Lagrange multiplier. This
generalizes the variational framework developed in \cite{GuvVaz}. The advantage of this approach is
that the multiplier quantifies directly the loss of Euclidean invariance associated with
the constraint;  it becomes a source for the stress induced in the surface, its value at any point
identified with the local normal force associated with the constraint.
\vskip1pc \noindent
If a surface of area greater than $4\pi$ is confined
within a unit hypersphere it will generally buckle into a
non-spherical shape. Unlike its loop counterpart,
however, this problem is not tractable analytically.  In this paper, it will be treated
perturbatively in the regime where the surface area is marginally greater than $4\pi$. In analogy to
a circular loop confined within a sphere,  it is found that the confined equilibrium shapes are
quantized.  These states are represented by spherical harmonic deformations of a two-sphere.  The
ground state is a quadrupole, exhibiting a three-fold degeneracy; all higher multipoles turn out to
be unstable.
\vskip1pc \noindent
The paper is organized as follows: In section \ref {confinedelastica}, the variational framework is
introduced. The Euler-Lagrange equations for the surface are derived and the constraining force
identified. Whereas the local confining force typically is
not constant, it will be if the confinement is within a hypersphere.
In section \ref{Spherical} the weak confinement of a two dimensional sphere by a
spherical hypersurface is examined. The paper concludes with a brief discussion and suggestions for
future work in section \ref{Discussion}. A number of useful definitions, identities and derivations
are collected in a set of appendices.

\section{\bf Confining surfaces by hypersurfaces} \label{confinedelastica}

Consider a surface $\Gamma_{N}$ of fixed area constrained to lie on a hypersurface $\Sigma_{N+1}$ in
an $N+2$-dimensional Euclidean space \cite{Spivak}.\footnote{Points here are represented ${\bf
x}=x^i \hat{\bf x}_i$, with $\hat{\bf x}_i$ the orthonormal basis in $\mathbb{R}^{N+2}$, ($i=1,
\dots, N+2$).} The surface and hypersurface embeddings in Euclidean space are described
parametrically by the maps $\Gamma_{N}: (s^1,\cdots,s^N) \rightarrow \mathbf{Y}(s^1,\cdots, s^N)$
and $\Sigma_{N+1}: (u^1,\cdots,u^{N+1}) \rightarrow \mathbf{X}(u^1,\cdots,u^{N+1})$ respectively.
The embedding of the surface on the hypersurface is described by a third mapping $\Gamma_{N,N+1}:
(s^1,\cdots,s^N) \rightarrow (U^1(s^1,\cdots,s^N),\cdots,U^{N+1}(s^1,\cdots,s^N))$. The three
mappings are related by composition, ${\bf Y} = {\bf X} \circ U$. While the problem of specific
interest will be in a two-dimensional surface embedded in a three-dimensional hypersphere, the
dependence of the confinement process on the dimension as well as the confining geometry is not
without interest and involves no significant additional calculational cost.
\vskip1pc \noindent
If the surface is freely deformable its bending energy is given by \cite{Polyakov}
\begin{equation} \label{Hamk2}
H = \frac{1}{2}\int dA_N \, (\nabla^2 {\bf Y})^2 \,,
\end{equation}
where $\nabla^2$ is the Laplacian on $\Gamma_N$ and $A_N$ its area, both of which are constructed
using the induced metric on $\Gamma_N$, $g_{ab}$, as described in the appendix.\footnote{It is
more usual to express $H$ in terms of curvatures. This is done explicitly in \ref{appembident}. For
surfaces of dimension higher than two, $H$ is not the most general bending energy of a closed surface. For example, if $N\ge 3$, there is an additional ``Einstein-Hilbert'' bending energy, proportional to the integrated scalar curvature. While this energy is topological when $N=2$ it is not for higher dimensional surfaces.}
\vskip1pc \noindent
In order to enforce the condition that $\Gamma_N$ lies on $\Sigma_{N+1}$, a term enforcing this constraint
is added to the energy $H[\mathbf{Y}]$ given by Eq. (\ref{Hamk2}):
\begin{equation} \label{confconst}
H_c [\mathbf{Y}, U] = H[\mathbf{Y}] + \int \, dA_N \,\,
\bm{\lambda}(s) \cdot \left[ \mathbf{Y}(s) - \mathbf{X} (U(s))
\right] + \mu \left(\int dA_N - {\cal A}_0\right) \,,
\end{equation}
which involves a vector-valued Lagrange multiplier $\bm{\lambda}$
defined along on $\Gamma_N$. This constraint will break the
manifest translational invariance of the energy $H$.
A second term enforces the constraint fixing the area.
It involves a Lagrange multiplier $\mu$ which is constant.
\vskip1pc \noindent
The response of $H_c$ to a change $\mathbf{Y}\to {\bf Y}+ \delta {\bf Y}$
in the embedding functions can be cast in the form
\begin{equation}
\label{delHc}
\delta_{\bf Y} H_c = \int dA_N\,
 \left(\nabla_a \mathbf{f}^a + {\bm \lambda}\right) \cdot
\delta \mathbf{Y}\,,
\end{equation}
where $\nabla_a$ is the covariant derivative on $\Gamma_N$
compatible with the induced metric $g_{ab}$ and the surface stress
 ${\bf f}^a$ is  given by
\begin{equation} \label{eq:stressdef}
{\bf f}^a = {\bf f}^a_B - \mu g^{ab} {\bf e}_b \,;
\end{equation}
it involves a contribution associated with the bending energy given by \cite{Stress}
\begin{equation} \label{eq:stressdef0}
{\bf f}^a_B = K_I \left( K^{abI}- \frac{1}{2} g^{ab} K^I\right)\,
{\bf e}_b - \tilde\nabla^a K_I\, {\bf n}^I\,,
\end{equation}
as well as a tension associated with the constraint on the area.
The simplest derivation of Eq. (\ref{delHc}) involves a
straightforward generalization of the method of auxiliary
variables introduced in \cite{Aux}.  The details are provided in
\ref{derivationfa}. Here ${\bf e}_a = \partial {\bf Y}/\partial
s^a$, $a=1,\cdots,N$ are the tangent vectors to the surface
adapted to the parametrization; the corresponding induced metric
is given by $g_{ab}= {\bf e}_a\cdot {\bf e}_b$. The vectors ${\bf
n}_I$, $I=1,2$ are any two mutually orthogonal unit vectors into
$\mathbb{R}^{N+2}$ normal to $\Gamma_N$ and, for each value of
$I$, $K_{ab}^{\phantom{ab}I}= {\bf e}_a\cdot \partial_b {\bf n}^I$
is the extrinsic curvature (a symmetric surface tensor) associated
with the rotation of the normal ${\bf n}^I$ towards the surface as
parameter curves are followed along the surface. $K^I= g^{ab}
K_{ab}^{\phantom{ab}I}$ is its trace.\footnote{Note the identity
for the energy density $(\nabla^2 {\bf Y})^2 = K^I K_I$.} Surface
indexes are lowered and raised with the metric tensor and its
inverse, $g^{ab}$; normal indexes are lowered and raised with the
Kronecker delta. The covariant derivative defined by $\tilde
\nabla_a K^I = (\delta^{I}_{\phantom{I}J} \nabla_a +
\omega_{a\phantom{I}J}^{\phantom{a}I}) K^J$, involves the normal
connection $\omega_{a}^{\phantom{a}IJ}$ defined by
$\omega_{a}^{\phantom{a}IJ}= {\bf n}^I\cdot\partial_a {\bf n}^J=-
\omega_{a}^{\phantom{a}JI}$, associated with the freedom to
rotate the
normals among themselves. Thus the analogue of the Gauss equations
for the normals is $\partial_a {\bf n}^I = K_a^{\phantom{a}bI}{\bf
e}_b-\omega_{a\phantom{a}J}^{\phantom{a}I}{\bf n}^J$.
This description is completely invariant with respect to local
rotations of the normal vectors. Confinement will select the
normal to the hypersurface as one of these normals, thus breaking
this invariance.
\vskip1pc \noindent In equilibrium, Eq. (\ref{delHc}) implies that
\begin{equation}
\nabla_a \mathbf{f}^a=-\bm{\lambda}\,.
\end{equation}
Thus, in the presence of the constraint, the stress in the surface is not conserved; the
Lagrange multiplier $\bm{\lambda}$ is identified as the external force associated with the
constraint \cite{LandauLifshitz}.
\vskip1pc \noindent
The corresponding variation of $H_c$ with respect to $U^A(s)$ is given by
\begin{equation}
\delta_U H_c = - \int dA_N \,\bm{\lambda} \cdot \mathbf{E}_A\, \delta U^A\,,
\end{equation}
where ${\bf E}_A =\partial {\bf X}/\partial u^A$, $A=1,\cdots,N+1$
are the $N+1$ tangent vectors to the hypersurface $\Gamma_N$
adapted to the parametrization defined by the coordinates $u^A$.
In equilibrium, $\bm{\lambda}\cdot {\bf E}_A =0$, and thus the
force on the curve associated with the contact constraint always
acts orthogonally to the surface.  One thus can write
$\bm{\lambda} = \lambda \,\mathbf{n}$, where $\mathbf{n}$ is the
unit vector normal to the hypersurface, $\Sigma_{N+1}$, and thus
\begin{equation} \label{EL}
\nabla_a \mathbf{f}^a = - \lambda \mathbf{n}\,.
\end{equation}
Using Eq. (\ref{eq:stressdef}), a straightforward calculation implies that the divergence
of ${\bf f}^a$ is normal to the surface and may be decomposed accordingly:
\begin{equation}\label{eq:FprimeEL}
\nabla_a \mathbf{f}^a  = \varepsilon_I \mathbf{n}^I\,,
\end{equation}
where the Euler-Lagrange derivatives $\varepsilon_I$ of the bending energy along the normal
directions ${\bf n}^I$, are  given by
\begin{equation} \label{shapeequation}
\varepsilon_I  = -\,\tilde \nabla^2 K_I + \frac{1}{2} (K_I K_J - 2 K_{I\, ab} K_J^{ab}) K^J +  \mu
K_I \,,
\end{equation}
where $\tilde{\nabla}^2=g^{ab} \tilde{\nabla}_a \tilde{\nabla}_b$. The two tangential Euler-Lagrange
derivatives $\nabla_a \mathbf{f}^a \cdot \mathbf{e}_b$ vanish identically, a consequence of the fact
that the Hamiltonian depends only on the surface geometry so that tangent deformations get
identified with reparametrizations.
\vskip1pc \noindent
The surface normals can always be rotated so that one of the two coincides with the hypersurface
normal ${\bf n}$. The second normal (${\bf l}$) is then uniquely fixed (modulo a sign) by the
conditions, ${\bf l}\cdot{\bf n}=0$, and ${\bf l}\cdot {\bf e}_a =0$.\footnote{An arbitrary basis
of normal vectors is reconstructed from the normal basis $\{{\bf n}, {\bf l}\}$ by a local rotation:
\begin{equation}
{\bf n}_1= \cos\Omega\, {\bf n} + \sin\Omega  \,{\bf l}\,;\quad
{\bf n}_2= - \sin\Omega \, {\bf n} + \cos\Omega\, {\bf l}\,.
\end{equation}}
With the choice ${\bf n}_1={\bf n}$ and ${\bf n}_2={\bf l}$ it is possible to write
\begin{equation} \label{eq:nabfn}
\nabla_a \mathbf{f}^a = \varepsilon_{\bf n} \, \mathbf{n} + \varepsilon_{\bf l}\, \mathbf{l} \,.
\end{equation}
Comparing Eq.(\ref{eq:nabfn}) with Eq. (\ref{EL}),  the Euler-Lagrange equation
\begin{equation} \label{eulelsurfeneb}
\varepsilon_\mathbf{l} = 0
\end{equation}
is identified.
The corresponding projection onto ${\bf n}$ determines the normal confining  force $\lambda$:
\begin{equation} \label{magnlambda}
\lambda =  - \varepsilon_{\bf n} \,.
\end{equation}
Both the Euler-Lagrange equation (\ref{eulelsurfeneb}) and the confining force (\ref{magnlambda}) are constructed out of the two unconstrained Euler-Lagrange derivatives. While the latter is completely determined by the local geometry, what this local geometry is will, of course, depend on the global behavior of the confined surface.
\vskip1pc \noindent
Eqs. (\ref{eulelsurfeneb}) and (\ref{magnlambda}) are still not very useful in their present
form. To facilitate their interpretation as well as their application, it is useful to cast
them explicitly with respect to variables adapted to the environment of the fixed confining
hypersurface.
\vskip1pc \noindent
Let $\kappa_{ab}$ represent the extrinsic curvature of $\Gamma_{N,N+1}$ embedded in $\Sigma_{N+1}$
(associated with the rotation of the normal ${\bf l}$), and $K_{AB}$ the extrinsic curvature of
$\Sigma_{N+1}$ embedded in $\mathbb{R}^{N+2}$ (associated with the rotation of its unique
normal ${\bf n}$) as described in \ref{appembident}.
\vskip1pc \noindent
In the adapted frame, the normal connection $\omega_{a\, 12}$ is identified as a surface vector,
\begin{equation}
\omega_a := \omega_{a12} = -{\cal E}_a^A l^B K_{AB}\,,
\end{equation}
where ${\cal E}^A_a$ are the components of the tangent vector ${\bf e}_a$ with respect to the basis
of tangent vectors to the hypersurface $\{{\bf E}_A\}$, ${\bf e}_a = {\cal E}^A_a{\bf E}_A$
with ${\cal E}^A_a = \partial U^A/\partial s^a$, and $l^A$ are the components of the vector normal
to $\Gamma_N$, tangent to $\Sigma_{N+1}$, ${\bf l}=l^A {\bf E}_A$.
\vskip1pc\noindent
Straightforward calculations give (details are presented in
\ref{appELder})
\begin{subequations}
\begin{eqnarray}
\varepsilon_{\bf l}  &=& - \nabla^2\kappa + \kappa \left(\frac{\kappa^2}{2} - \kappa_{ab} \kappa
^{ab}+ \mu\right) + K_{(1)}\left(\frac{\kappa K^{(1)}}{2} - \kappa_{ab} K^{ab\,(1)}\right)
+ \Omega_{\bf l}\,;\label{Euler1}\\
\varepsilon_{\bf n} &=& - \nabla^2 K^{(1)} + \kappa \left(\frac{\kappa K_{(1)}}{2}-
\kappa_{ab} K^{ab\,(1)}\right) + K_{(1)}\left(\frac{K_{(1)}^2}{2} - K_{ab
\,(1)} K^{ab\,(1)} +\mu \right) + \Omega_{\bf n}\,, \label{Euler2}
\end{eqnarray}
\end{subequations}
where $K_{(1)\,ab}$ is identified with the projection of $K_{AB}$ onto $\Gamma_{N,N+1}$, and thus
given by (see \ref{appembident})
\begin{equation} \label{eq:K1ab}
K_{(1)\,ab}= {\cal E}^A_a {\cal E}^B_b K_{AB} \,.
\end{equation}
$\Omega_{\bf l}$ and $\Omega_{\bf n}$ are given by
\begin{subequations}
\begin{eqnarray}
\Omega_{\bf l} &=&  K^{(1)}\nabla \cdot \omega + 2 \omega\cdot
\nabla K^{(1)} + |\omega |^2 \kappa\, \label{omegal}\,;\\
\Omega_{\bf n} &=& - \kappa \nabla \cdot \omega - 2 \omega \cdot
\nabla \kappa + |\omega|^2  K_{(1)}\,. \label{omegan}
\end{eqnarray}
\end{subequations}
They both vanish when $\omega_a$ does. The divergence and squared
norm of $\omega_a$ are given by
\begin{subequations}
\begin{eqnarray}
\nabla\cdot \omega &=& \Big(\kappa l^A l^B - {\cal H}^{AB} l^C\nabla_C\Big)K_{AB}-\kappa_{ab}
K^{ab\,(1)}\,;\\
|\omega |^2 &=& l^Al^B \left(K^{(1)}K_{AB} - R_{AB}\right)\,,
\end{eqnarray}
\end{subequations}
 where ${\cal H}^{AB}  =  g^{ab} {\cal E}^A_a {\cal E}^B_b = G^{AB}- l^A l^B$ is the projector onto
$\Gamma_{N,N+1}$ and $R_{AB}$ is the Ricci tensor on the hypersurface.

\subsection{Confinement of an elastic curve by a surface}

Before examining surfaces confined within a hypersphere, it is useful to confirm that
Eqs.(\ref{Euler1}) and (\ref{Euler2}) reproduce the expressions describing the confinement of
elastic curves by a two-dimensional surface in three-dimensional Euclidean space derived in Ref.
\cite{GuvVaz}.
\vskip1pc \noindent
Let the curve be parametrized by arc-length $s$. Its tangent ${\bf e}_s={\bf T}$  is now a unit
vector so that $g_{ss}=1$; $\nabla_s$ is the derivative with respect to $s$, denoted by a prime
$'$. The extrinsic curvature tensors for each of the two normals are scalars along the curve: one
identifies $\kappa_{ss} = \kappa = \kappa_g$ as the geodesic curvature; the projection $K^{(1)}_{ss}
= K^{(1)} = \kappa_n$ as the normal curvature; and the normal connection $\omega_{s12}=-t^A l^B
K_{AB}=\tau_g$ as the geodesic torsion. On account of these identities, the normal covariant
derivatives assume the form $\tilde{\nabla}^s K^{(1)}=\kappa'_n + \kappa_g \tau_g$ and
$\tilde{\nabla}^s \kappa = \kappa'_g - \kappa_n \tau_g$. The stress tensor is now represented by a
vector ${\bf F}$ along the curve, defined by
\begin{equation}
{\bf F} := {\bf f}^s= \left(\frac{\kappa^2_g+\kappa^2_n}{2}-\mu\right) {\bf T}-\left(\kappa'_n +
\kappa_g \tau_g\right){\bf n} - \left(\kappa'_g - \kappa_n \tau_g\right){\bf l}\,.
\end{equation}
In addition,
$\nabla \cdot \omega = \tau'_g$ and $|\omega|^2=\tau^2_g$, so that Eqs.(\ref{Euler1}) and
(\ref{Euler2}) reduce to
\begin{subequations}
\begin{eqnarray}
-\varepsilon_{\bf l} &=& -{\bf F}' \cdot {\bf l} = \kappa_g'' + \kappa_g \left(\frac{\kappa_g^2 +
\kappa_n^2}{2} - \tau_g^2 - \mu\right) -\frac{ (\kappa_n^2\tau_g)'}{\kappa_n} \,,\label{Curve1}\\
-\varepsilon_{\bf n} &=& -{\bf F}' \cdot {\bf n} = \kappa_n'' + \kappa_n \left( \frac{\kappa_g^2 +
\kappa_n^2}{2} - \tau_g^2 - \mu\right) +\frac{ (\kappa_g^2\tau_g)'}{\kappa_g}\,.\label{Curve2}
\end{eqnarray}
\end{subequations}
Setting $\varepsilon_{\bf l}=0$ reproduces the Euler-Lagrange
equation describing a confined elastic curve obtained by Nickerson
and Manning \cite{Mann, MannNick}. Eq. (\ref{Curve2}) identifies
the magnitude of the normal force written down in \cite{GuvVaz}.
\vskip1pc \noindent If the curve is confined by a sphere (with
$\kappa_n=1$ and $\tau_g=0$), its equilibrium states are described
by the Euler-Lagrange equation, $-\varepsilon_{\bf
l}=\kappa''_g+\kappa_g(\kappa^2_g/2-\mu+1/2)$ \cite{LangSing}. The corresponding
normal force simplifies to $\lambda=-\varepsilon_{\bf n} =
\kappa_g^2/2-\mu+1/2$. A detailed treatment of confinement of
elastic curves by spheres was presented in \cite{GuvVaz}. In this
case, as shown in \cite{GuvVaz}, it is possible to exploit the
residual rotational invariance of the problem. This implies that
the torque vector ${\bf M}$ is conserved, where ${\bf M}$ is given
by ${\bf M}  = {\bf X} \times {\bf F}+{\bf S}$, a sum of the
moment of the force ${\bf F}$ and an intrinsic moment ${\bf S} =
-\kappa_g {\bf n} + \kappa_ n {\bf l}$.\footnote {${\bf M}$ is the
one-dimensional reduction of the surface torque tensor, ${\bf
m}^{ai} = \varepsilon^{ijk} X^j f^{ak}+S^{ai}$, where $S^{ai} =
K^I \varepsilon^{ijk} e^{aj} n^{Ik}$ discussed in
\ref{approtation}.} The equation ${\bf M}^2 =$ constant provides a
quadrature for $\kappa_g$.
The loop is then constructed directly from its curvature data
using  a polar chart adapted to the vector ${\bf M}$.

\section{Confinement by hyperspheres}
\label{Spherical}

\subsection{Hyperspherical shape equation and transmitted force}

If the hypersurface is a unit $N+1$-sphere, its extrinsic
curvature is given by $K_{AB}=G_{AB}$ so that
$K=N+1=K^{AB}K_{AB}$.  Using Eq.(\ref{eq:K1ab}), one then
identifies $K^{(1)}_{ab}=g_{ab}$ and therefore $K^{(1)}=N$. In
addition, $\omega_a=-{\cal E}^A_a l_A=0$, so that $\Omega_{\bf
l}=0=\Omega_{\bf n}$. \vskip1pc \noindent The stress tensor is
given by ${\bf f}^a=f^{ab}{\bf e}_b+f^a {\bf l}$, where
\begin{subequations}
\begin{eqnarray}
f^{ab} &=& \kappa \left(\kappa^{ab}-\frac{1}{2}\kappa g^{ab}
\right) - \left(\frac{N}{2}(N-2)+\mu\right)g^{ab}\,; \\
f^a  &=& -\nabla^a \kappa
\end{eqnarray}
\end{subequations}
The spherical Euler-Lagrange equation is then given by
\begin{equation} \label{elsph}
\varepsilon_{\bf l}  = \nabla_a {\bf f}^a \cdot {\bf l} = -\nabla^2 \kappa +
\kappa\left(\frac{\kappa^2}{2}-
\kappa_{ab}\kappa^{ab} + \frac{N}{2} (N-2) + \mu \right) =0\,.
\end{equation}
The magnitude of the corresponding normal force is
\begin{equation} \label{forcesph}
-\lambda = \varepsilon_{\bf n} = \nabla_a {\bf f}^a \cdot {\bf n}= \frac{\kappa^2+N^2}{2}(N-2)+\mu
N\,.
\end{equation}
These are the hyperspherical reductions of Eqs. (\ref{Euler1}) and
(\ref{Euler2}). An alternative derivation of Eq.(\ref{elsph}) is
provided in \ref{apphypbend} which exploits the hyperspherical
geometry in the variational principle. Notice, however, that the
derivation presented there suffers the disadvantage of having
nothing to say about the normal forces described by
Eq.(\ref{forcesph}).  This involves the behavior of the surface
energy under deformations normal to the hypersurface and thus
falls beyond the scope of any derivation that is intrinsic to the
hypersurface.
\vskip1pc \noindent
When $N=2$, the force $\lambda$ is completely determined by the multiplier $\mu$, $\varepsilon_{\bf n}=2\mu$. It
is thus constant along the contact. This is not true if the confining surface is not spherical. In the next section, the
origin of this coincidence will be traced to the scaling behavior of the constrained energy.

\subsection{Virial identities} \label{sect:virial}

The bending energy of a two-dimensional surface
has one particularly distinctive feature: its scale invariance.
It is also invariant under the conformal transformations of the
surfaces induced by embedding from its environment
\cite{Willmore}. A consequence of scale invariance is that tension is not introduced in the surface when its area is fixed unless additional constraints are introduced to set a scale;  a circular loop, on the other hand,  would dilate
if its length  were not constrained; a three or higher
dimensional surface would collapse.
This behavior reflects the dimensional dependence of the behavior of
bending energy under scaling. Confinement will break the
scale invariance of the two-dimensional bending energy.
\vskip1pc\noindent
More specifically, the behavior under scaling of the constrained energy in equilibrium implies the identity
\begin{equation} \label{eq:lambdamu}
\int dA_N \, {\bf n}\cdot {\bf Y}\, \lambda  =
-(N-2)\, H[\mathbf{Y}] - \mu N A_N \,,
\end{equation}
connecting a weighted surface averaged $\lambda$, the bending
energy $H$, and the multiplier $\mu$. When $N=2$, there is no
explicit dependence on $H$. To derive Eq. (\ref{eq:lambdamu}),
consider the effect of a rescaling ${\bf Y}\to \Lambda {\bf Y}$ on
the constrained energy $H_c$, given by Eq.(\ref{confconst}). This
gives
\begin{equation} \label{Hscale}
H_c [\Lambda \mathbf{Y}, U] = \Lambda^{N-2} H[\mathbf{Y}] +
\Lambda^N \int \, dA_N \,\, \bm{\lambda}(s) \cdot \left[\Lambda
\mathbf{Y}(s) - \mathbf{X} (U(s)) \right] + \mu (\Lambda^N A_N -
{\cal A}_0)\,,
\end{equation}
so that in equilibrium,
\begin{equation}
 \frac{d H_c [\Lambda \mathbf{Y}, U]}{d\Lambda}\Big|_{\Lambda=1}= 0\,.
\end{equation}
This reproduces the identity Eq. (\ref{eq:lambdamu}). \vskip1pc
\noindent In the case of interest, $N=2$, the bending energy is
scale invariant;  it  does not appear explicitly in
Eq. (\ref{eq:lambdamu}). \vskip1pc \noindent
 If the surface $\Gamma_N$ is closed, Eq. (\ref{EL}) implies
 the integrability condition
\begin{equation}
\label{intlambda} \int dA_N \, \lambda\, {\bf n}=0\,.
\end{equation}
Contact with the hypersurface does not need to be complete. This
identity also follows from the translation invariance of Eq.
(\ref{eq:lambdamu}). More explicitly, let ${\bf Y}\to {\bf Y} +
{\bf a}$, where ${\bf a}$ is a constant vector. The only
non-invariant term in Eq. (\ref{eq:lambdamu}) is the first one.
\vskip1pc \noindent If $\Sigma_{N+1}$ is a unit hypersphere, Eq.
(\ref{eq:lambdamu}) implies the remarkable formula
\begin{equation} \label{eq:Pi}
\Pi := \int dA_N \, \lambda= - (N-2)\, H[\mathbf{Y}] - \mu N A_N\,,
\end{equation}
connecting the total force transmitted to the hypersphere to
bending energy, the multiplier $\mu$ and the area.  While it is also
implied directly by Eq. (\ref{forcesph}), its derivation
as an identity associated with scaling presented here does not require
any local input, and as such is more compelling.
There are two points worth emphasizing:
\vskip1pc \noindent
(i)  In the absence of a constraint on the area, there will be no force on the containing hypersphere: $\Pi$
is completely determined by the product $\mu A_2$. Indeed
$\lambda$ itself is constant, proportional to $\mu$. This is a direct
consequence of the scale invariance of the bending energy of a
two-dimensional surface.  In striking
contrast with the behavior of a confined circular loop. the
confining force is due entirely to the area constraint and is
constant over the surface.
\vskip1pc \noindent (ii) The contribution of bending energy to the
transmitted force in Eq. (\ref{eq:Pi}) changes sign when $N=2$.
This reflects the contrasting behavior of a loop ($N=1$), and a
three or higher dimensional surface. In the former case, the area
(length) constraint works against the tendency of bending energy
to expand the loop; whereas in the latter it prevents its
collapse.

\subsection{Weakly Confined  Spheres} \label{WeakSpherical}

\begin{figure}[htb]
\begin{center}
 \includegraphics[scale=0.6]{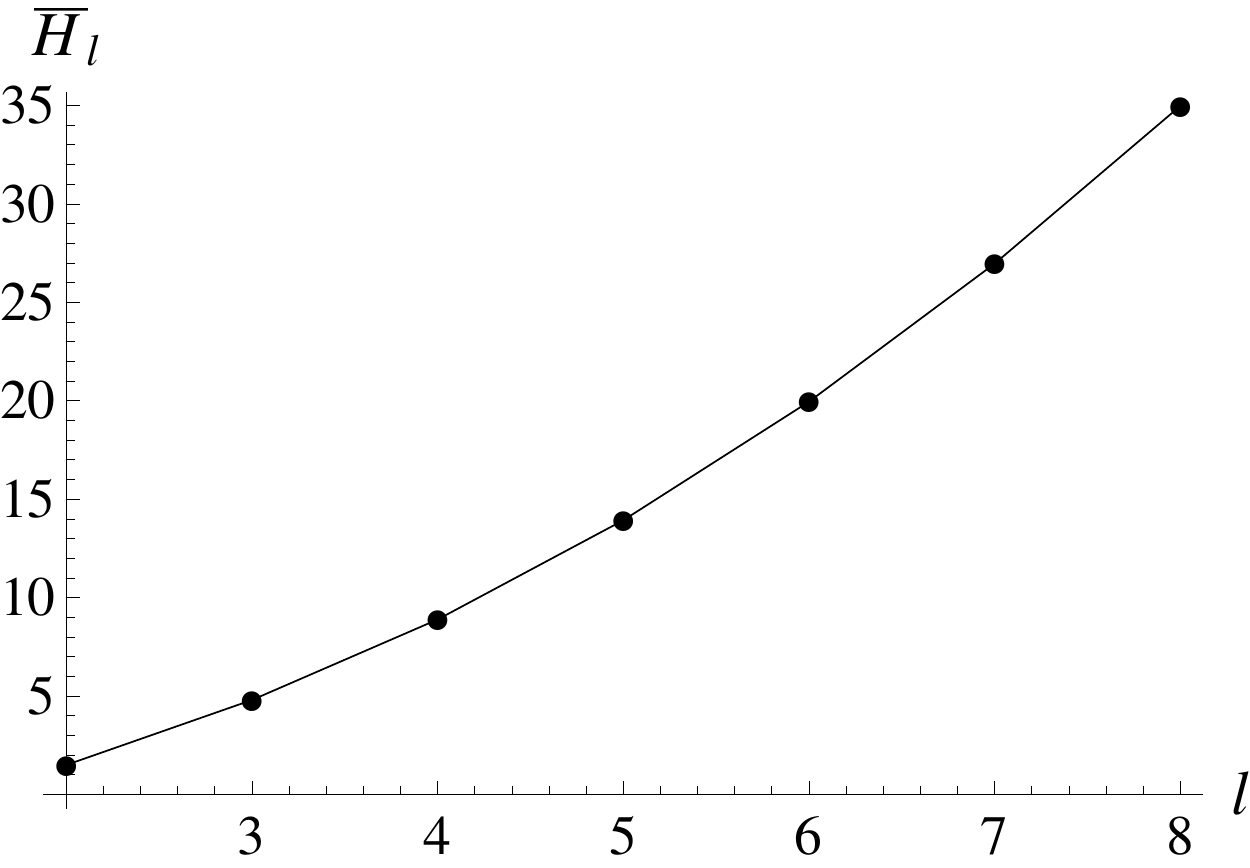}
\end{center}
\caption{\small{Energy per excess area of the deformed stated for  low values of $l$}}
\label{fig2}
\end{figure}

\begin{figure}[htb]
\begin{center}
\includegraphics[scale=0.15]{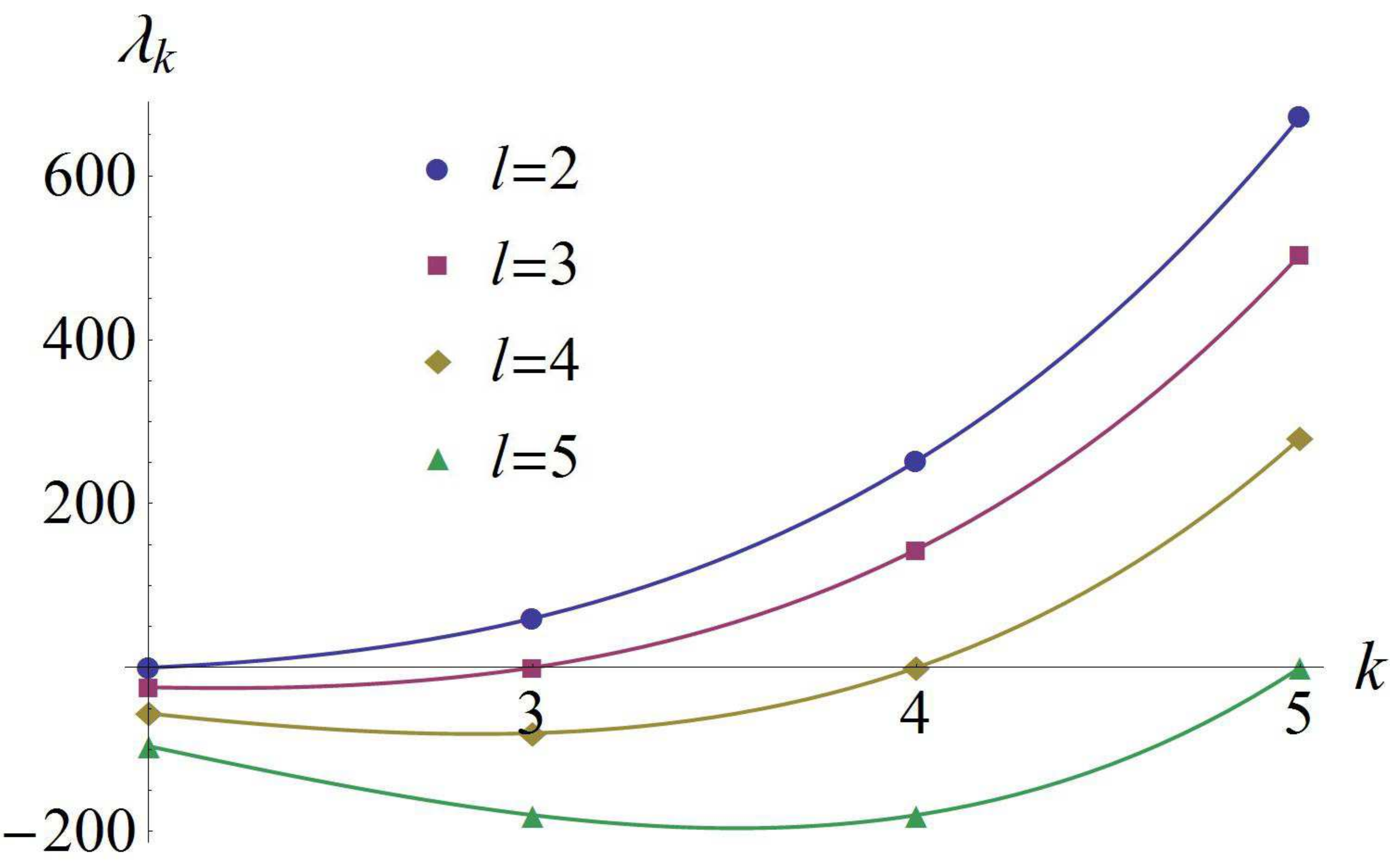}
\end{center}
\caption{\small{Eigenvalues of ${\cal L}$ for low values of $l$. Each point corresponds to $2 k+1$ deformation modes on account of the spherical harmonics degeneracy}}
\label{fig3}
\end{figure}

Consider a surface of area ${\cal A}_0/4\pi -1 \ll 1$ so that it
is weakly confined within a unit hypersphere. In the linear
approximation, the surface  can be described in terms of a
deformation of a two-sphere of unit radius (${\bf X}_0$ say) along
the normal ${\bf l}$ into the hypersphere. The surface can thus be
characterized by the amplitude of the deformation  $\Phi$,
\begin{equation}
{\bf X}= {\bf X}_0 + \Phi\, {\bf l}\,.
\end{equation}
More generally, consider the deformation of a surface with metric
tensor $g_{ab}$ and extrinsic curvature $\kappa_{ab}$ embedded in
a Riemannian manifold with metric $G_{AB}$. The metric and
extrinsic curvature of the surface deformed along the normal ${\bf
l}$ are given by $\delta g_{ab}= 2 \kappa_{ab} \Phi$, and
\begin{equation}
\delta \kappa_{ab}= -
\nabla_a\nabla_b \Phi + [\kappa_{ac}\kappa^{c}_{\phantom{c}b} - R_{ABCD} l^A
l^C {\cal
E}^ B_a {\cal E}^D_b ] \Phi\,,
\end{equation}
where $R_{ABCD}$ is the Riemann tensor constructed with $G_{AB}$
(see, for example, \cite{CapGuv} for a derivation). Thus
 \begin{equation}  \label{delkap}
\delta \kappa = - \nabla^2 \Phi - [\kappa_{ab}\kappa^{ab} + R_{AB}
l^A l^B ] \Phi\,.
\end{equation}
The sphere is totally geodesically embedded in the hypersphere, so that $\kappa_{ab} = 0$.
For a deformed unit sphere, Eq. (\ref{delkap}) thus reads
 \begin{equation}
\delta \kappa =- [\nabla_0^2   + 2] \Phi \,,
\end{equation}
where $\nabla_0^2$ is the Laplacian on the unit two-sphere. There
is no corresponding first order change in the induced metric. The
linearization of Eq. (\ref{elsph}) is thus given by
\begin{equation} \label{eq:linEL}
[  \nabla_0^2  + 2  +  \mu ] [\nabla_0^2 + 2] \Phi =0\,.
\end{equation}
$\Phi$ is now expanded in spherical harmonics,
 \begin{equation}
\Phi = \sum_{l\ne 0, m} \alpha_{lm} Y_{lm} \,.
 \end{equation}
There are three zero modes with $l=1$, which correspond to a rotated two-sphere. The mode with $l=0$ corresponds to a
dilatation of the sphere changing its area so is excluded by the area constraint. Global solutions of Eq.(\ref{eq:linEL}) imply a fixed single value of $l$. The linearized equilibrium states are thus spherical harmonic deformations of the two-sphere. This result is also directly analogous to the behavior of an elastic curve confined by a sphere \cite{GuvVaz}. The force per unit area transmitted to the hypersphere is given, for each $l$, by
\begin{equation} \label{mul}
\mu_l = l(l+1)-2\,.
\end{equation}
It is constant and independent of the magnitude of $\Phi$. The latter observation may appear to be counter-intuitive: it can, however, be understood as a higher dimensional analogue of the Euler buckling instability in an unstretchable rod
\cite{LandauLifshitz}. A given critical force will be necessary to buckle the sphere into each harmonic.  The total force does, however, depend on the excess area.
\vskip1pc \noindent
To determine the energy of the deformed sphere, note that the excess area is related to the deformation as
follows \footnote{The first order change in area, given by $\delta A= \int d\Omega_2 \kappa \Phi$, vanishes \cite{Spivak,CapGuv}.}
\begin{equation}
A \approx 4\pi + \int d\Omega_2\, \delta \kappa \Phi= 4\pi - \int d\Omega_2\, \Phi [ \nabla_0^2 + 2 ]\Phi\,,
\end{equation}
where $d\Omega_2$ is the area measure on the unit two-sphere. Thus
\begin{equation}
A - 4\pi = \mu_l \, \sum_m  \, |\alpha_{lm}|^2\,.
 \end{equation}
The energy of the deformed state with a fixed value of $l$ is then given by \footnote{This is obtained most directly by
expanding Eq. (\ref{Hamkappa}) with $N=2$ about a unit two-sphere.}
\begin{eqnarray} \label{sphenergy}
H_l &\approx& \frac{1}{2}\, \int d\Omega_2\, [ \delta \kappa^2 +  4 \delta \kappa \Phi]\nonumber\\
  &=& \frac{1}{2}\, (\mu_l - 2) (\mu_l + 2) \sum_m \,|\alpha_{lm}|^2\nonumber\\
  &=& \frac{1}{2}\, \frac{(\mu_l - 2) (\mu_l + 2)}{\mu_l}\, (A- 4\pi) \,,
\end{eqnarray}
where $\mu_l$ is given  by Eq(\ref{mul}). It increases linearly with area. Energy per unit excess area, $\bar{H}_l=H_l/(A-4 \pi)$, is plotted as a function of
$l$ in Fig. \ref{fig2}; it increases monotonically with $l$.
\vskip0pc \noindent
The ground states are represented by ellipsoidal (or quadrupole) deformations with
$l=2$. Modulo its orientation with respect to the sphere, the
ground state is three fold degenerate.\footnote {Note that, at
this order,  prolate and oblate deformations of the sphere are
interchanged under a sign change in $\Phi$.} One would expect this
degeneracy to be lifted at higher orders in the perturbation.
\vskip1pc \noindent
Each of these perturbative states of the confined sphere will possess analogues in the full non-linear
description. In particular, an infinite number of axially symmetric states solutions, one for each $l$, would be
expected; ellipsoids, pears, and dumbbells (or discocytes) corresponding to $l =2,3,4$ respectively. A detailed analysis
is beyond the scope of this paper but will be taken up elsewhere \cite{GuvVaz1}. It would be interesting to know if cross talk between different values of $l$ emerges as non-perturbative features of the confinement process giving rise to composite structures such as (cup-shaped) stomatocytes.
\vskip1pc \noindent
The analysis of the confined elastic loop suggests that the simple dependence of the
energy and the transmitted forces on the excess area will diverge from the
simple behavior described here as the confined surface explores more of the hyperspherical volume and deviations
from geodesic behavior become increasing pronounced.
\vskip1pc \noindent
In analogy with an elastic loop, it is reasonable to expect that the only stable weakly confined states
are the ellipsoidal ground states with $l=2$.  All excited states (with $l\ge 3$)  are unstable.  To show this,  note that the second variation of the energy about a weakly confined state is given by
\begin{equation} \label{2ndvar}
H_2 \approx \frac{1}{2}\, \int d\Omega_2\, \Phi {\cal L}_l \Phi\,,
\end{equation}
where the operator ${\cal L}_l $ is defined by
\begin{equation}
{\cal L}_l= [\nabla_0^2  + l(l+1) ] [\nabla_0^2 + 2] \,.
\end{equation}
The eigenvalues of ${\cal L}_l$ are identified as
\begin{equation}
\lambda_k = (k(k+1) - l(l+1))(k(k+1) - 2) \,.
\end{equation}
These eigenvalues are plotted  for $l=2,3,4$ and $5$ in Fig. \ref{fig3}.
One observes that for each $l\geq2$ there will be $l^2-4$ negative eigenvalues corresponding to unstable modes of decay. Thus, as claimed previously, the only stable states are those with $l=2$. This generalizes the qualitatively very similar stability analysis of loops weakly confined within a sphere \cite{GuvVaz}. Whether this behavior is a faithful representation of the behavior when the surface is not weakly confined, on the other hand, is anyone's guess. Indeed, it will no longer be legitimate to assume that the ground state is completely attached.

\section{\bf Conclusions} \label{Discussion}

In this paper, the confinement of a freely deformable closed
two-dimensional surface  within a three-dimensional hypersphere
of smaller radius was examined.  This is the natural generalization of
a problem of a more domestic nature: the confinement of a loop within a surface.
In the absence of the constraint the unique non-singular--topologically spherical--equilibrium state is a
round sphere. Under confinement, this sphere deforms into one of an infinite number of equilibrium shapes; they have been
examined in the limit of weak confinement.  Mathematically, this problem can be viewed as a natural generalization of the spherical Willmore problem, to which it reduces when the constraint on the
area is relaxed. In this context, it is also reasonable to expect
that counterparts of these states exist in each topological sector
of the two-dimensional surface.
\vskip1pc \noindent
While the study of confinement in higher dimensions is not expected--or is unlikely--to offer insight into
problems that arise in biology or soft matter,  the questions
posed are of a fundamental nature that have not been explored
previously. Indeed, if the geometrical invariants of a surface are
ordered by derivatives, after area, the bending energy is arguably
the most important--non-topological--invariant among them.
Analogues of the problem could be relevant in relativistic field
theories, be it the Euclidean formulation of such theories, or the
behavior of  stationary spacelike surface states.  Analogous
actions were discussed by Polyakov as effective field theories
\cite{Polyakov}.
\vskip1pc \noindent
If the analysis of confined loops is  any guide, a non-perturbative extension of this work should be worth
pursuing. To this end,  it is possible to exploit the conformal
invariance of the two-dimensional bending energy \cite{Willmore}
to reformulate the problem in a rather striking way. For under
inversion in a hypersphere, itself located on the confining
hypersphere, the latter will get mapped to a three-dimensional
Euclidean hyperplane.  The confined surface on the three-sphere is
replaced by a surface that is free to bend in this
three-dimensional space without obstruction. The area, however, is
not invariant under conformal transformations; it gets replaced by
a surface integral
\begin{equation}
A\to \int\, \frac{ dA_2}{(|{\bf X}|^2 +\frac{1}{4})^2}\,\,,
\end{equation}
involving a  potential that will depend on the Euclidean distance $|{\bf X}|$ from points on the
surface to the origin of the three dimensional space.  The confining surface is replaced by this
potential.
\vskip1pc\noindent
It is also possible to consider confinement as a dynamical process in a relativistic
theory of extended objects. Two-dimensional surfaces are interpreted now as the world-sheets of strings \cite{Vilenkin,GSW}; the confining hypersphere is replaced by a three-dimensional de Sitter space,
the fixed worldsheet of a membrane.

\section*{Acknowledgements}

Partial support from DGAPA PAPIIT grant IN114510 as well as CONACyT grant 180901
is acknowledged. PVM is also grateful to UAM-Cuajimalpa for financial support.

\begin{appendix}

\setcounter{equation}{0}
\renewcommand{\thesection}{Appendix \Alph{section}}
\renewcommand{\thesubsection}{A. \arabic{subsection}}
\renewcommand{\theequation}{A.\arabic{equation}}

\section{Embedding identities} \label{appembident}

In this appendix, several useful identities connecting
$\Gamma_N,\Gamma_{N,N+1}$ and $\Sigma_{N+1}$, as defined in
section \ref{confinedelastica}, are collected. \vskip 1pc
\noindent Let ${\bf E}_A,$ $A=1,\dots N+1$ be the tangent vectors
to the hypersurface $\Sigma_{N+1}$, and ${\bf n}$ its normal
vector. The Gauss structure equations describing the embedding of
$\Sigma_{N+1}$ in Euclidean space are given by
\begin{equation}
\label{G3} \nabla_A {\bf E}_B = - K_{AB} {\bf n}\,.
\end{equation}
Here $\nabla_A$ is the covariant derivative on
$\Sigma_{N+1}$ compatible with its induced metric,
$G_{AB}= {\bf E}_A\cdot {\bf E}_B$.  These equations
define the curvature tensor $K_{AB}$.

\vskip1pc \noindent The counterpart of Eq.(\ref{G3}), describing
the embedding of the surface $\Gamma_{N}$ directly into the same
Euclidean space, is given by
\begin{equation}
\label{G4}
\nabla_a {\bf e}_b = - K_{ab\, I}\, {\bf n}^I\,.
\end{equation}
Choose ${\bf n}_1={\bf n}$, and ${\bf n}_2={\bf l}$. The surface
tangent vectors ${\bf e}_a$, $a=1,\cdots, N$ can be cast as a
linear combination of their hypersurface counterparts as follows,
${\bf e}_a = {\cal E}^A_a \, {\bf E}_A$, where ${\cal E}^A_a=
\partial U^A/ \partial s^a$. The vector ${\bf l}$ can likewise be
expanded, ${\bf l}= l^A {\bf E}_A$. \vskip1pc \noindent Projecting
$\nabla_A {\bf E}_B$ onto $\Gamma_{N}$ gives
\begin{equation}
{\cal E}^A_a {\cal E}^B_b \nabla_A {\bf E}_B = \nabla_a {\bf e}_b
-   (\nabla_a {\cal E}_b^B) \, {\bf E}_B\,.
\end{equation}
It follows from the structure equation (\ref{G3}) and its
definition, that the extrinsic curvature of
$\Gamma_N$ associated with the rotations of ${\bf n}$ onto
$\Gamma_N$ is  given by Eq.(\ref{eq:K1ab}),
$ K_{ab\,(1)}  = {\cal E}^A_a {\cal E}^B_b K_{AB}$.
In particular,  using the completeness of the basis vectors $\{{\cal E}^A_a,l^A\}$,
$g^{ab}{\cal E}^A_a {\cal E}^B_b+ l^A l^B = G^{AB}$,
the surface trace of $K_{(1)\,ab}$ is given by
\begin{equation} \label{K1}
 K_{(1)}  = g^{ab} {\cal E}^A_a {\cal E}^B_b K_{AB}=
(G^{AB}- l^A l^B) K_{AB} = K - l^A l^B K_{AB}\,.
\end{equation}
In addition,
\begin{equation}
{\cal E}^A_a l^B\nabla_A {\bf E}_B = \nabla_a {\bf l} -  (\nabla_a l^B) {\bf E}_B \,,
\end{equation}
so that
\begin{equation} \label{omega12}
\omega_{a\,12} := {\bf n} \cdot \nabla_a {\bf l} = - {\cal E}^A_a l^B K_{AB}\,.
\end{equation}
\vskip 1pc \noindent
If the surface is embedded in a hypersphere with $K_{AB}=  G_{AB}$, then $K_{(1)\,ab}=  g_{ab}$ and
$\omega_{a\,12} =0$.
\vskip1pc \noindent
For a curve, with $N=1$, with unit tangent vector ${\bf T}= t^A {\bf E}_A$,
$K_{(1)}= t^A t^B K_{AB} = \kappa_n$; $\omega_{12}=l^A t^B
K_{AB} = \tau_g$, where $\kappa_n$ and $\tau_g$ are the normal curvature and geodesic torsion of the
curve.
\vskip1pc \noindent
The counterpart of Eqs.(\ref{G3}) and (\ref{G4})  for $\Gamma_{N,N+1}$ embedded into $\Sigma_{N+1}$
are
\begin{equation} \label{G5}
D_a {\cal  E}_b^A  = - \kappa_{ab}\,  l^A\,,
\end{equation}
where
\begin{equation} \label{G55}
D_a {\cal E}_b^C := \nabla_a {\cal E}^C_b +\Gamma^{C}_{\phantom{C}AB} {\cal E}^A_a {\cal E}^B_b\,,
\end{equation}
and $\Gamma_{AB}^C$ is the Christoffel connection compatible with
the induced metric on $\Sigma_{N+1}$, $G_{AB}$. Eq.(\ref{G5})
defines the extrinsic curvature tensor $\kappa_{ab}$ associated
with the rotation of the normal ${\bf l}$ onto $\Gamma_N$.  This
provides the  identity for $K^{(2)}_{ab}$,
\begin{equation} \label{G23}
K^{(2)}_{ab}: = \kappa_{ab} = -l_B D_a {\cal E}^B_b \,.
\end{equation}

\setcounter{equation}{0}
\renewcommand{\thesection}{Appendix \Alph{section}}
\renewcommand{\thesubsection}{B. \arabic{subsection}}
\renewcommand{\theequation}{B. \arabic{equation}}

\section{Derivation of the stress tensor} \label{derivationfa}

In this appendix, the stress tensor introduced in Eq.(\ref{delHc}) will be derived. This will be
done by extending the auxiliary variables method introduced in \cite{Aux} to co-dimensions higher
than one. One constructs a
functional $H_c[{\bf X},{\bf e}_a, {\bf
n},g_{ab},K_{ab},\omega_a^{\phantom{a}IJ},{\bf
f}^a,\lambda^{a}_{\phantom{a}I}, \lambda_{IJ}, \lambda^{ab},
\Lambda^{ab}_{\phantom{ab}I}, \Lambda^ {a}_{ \phantom{a}IJ}]$
treating the embedding functions ${\bf Y}$, the adapted basis $\{{\bf e}_a,{\bf n}^I\}$, the metric
tensor $g_{ab}$, the extrinsic curvature tensor
$K^I_{ab}$, as well as the connection $\omega_{a}^{\phantom{a}IJ}$
as independent variables, by introducing Lagrange multipliers
${\bf f}^a, \lambda^{a}_{\phantom{a}I}, \lambda_{IJ},
\lambda^{ab}, \Lambda^{ab}_{\phantom{ab}I}, \Lambda^ {a}_{
\phantom{a}IJ}$ \footnote{The Christoffel connection
$\Gamma^{c}_{\phantom{c}ab}$ is constructed using $g_{ab}$ and its
derivatives; in contrast, the normal connection
$\omega_{a}^{\phantom{a}IJ}$ is not constructed from the normal
``metric'' $\delta^{IJ}$ so that, in general, it is necessary to
implement its definition in the variational principle.}
\begin{align}
 H_c & = \frac{1}{2} \int dA_N\,  K_I \, K^I + \int dA_N \left[{\bf f}^a\cdot ({\bf e}_a-
\partial_a\, {\bf Y})
+\lambda^{ab} (g_{ab}-{\bf e}_a\cdot {\bf e}_b)+\frac{1}{2} \lambda_{IJ} (\delta^{IJ}-{\bf n}^I
\cdot {\bf n}^J)\right] \nonumber \\
&+\int dA_N \left[\lambda^{a}_{\phantom{a}I}{\bf e}_a \cdot {\bf
n}^I+\Lambda^{ab}_{\phantom{ab}I}(K_{ab}^{\phantom{ab}I}-\partial_a {\bf n}^I \cdot
{\bf e}_b)+\Lambda^{a}_{\phantom{a}IJ}(\omega_{a}^{\phantom{a}IJ}-{\bf n}^I \cdot \partial_a {\bf
n}^J)\right]\,.
\end{align}
The bending energy now depends only on the independent variables
$g_{ab}$ and $K^I_{ab}$.
Lagrange multipliers with more than one index possess the same
symmetry as the quantities they multiply. Thus, for instance,
$\lambda^{ab}=\lambda^{ba}$, $\lambda_{IJ}=\lambda_{JI}$,
$\Lambda^ {a}_{\phantom{a}IJ}=-\Lambda^ {a}_{\phantom{a}JI}$. The
variation of $H$ with respect to ${\bf Y}$ gives
\begin{equation}
 \delta_{\bf X}H_C=\int dA_N\, \nabla_a{\bf f}^a \cdot \delta {\bf Y}\,.
\end{equation}
Thus, in equilibrium, ${\bf f}^a$ will be conserved on a free
surface; it is identified as the stress tensor on the surface
\cite{Aux}. The Euler-Lagrange equations for ${\bf e}_a$ provide
an expansion of the Lagrange multiplier ${\bf f}^a$ with respect
to the adapted basis
\begin{equation} \label{falambda}
{\bf f}^a = \left(2 \lambda^{ab}+\Lambda^{acI} K_{c\phantom{c}I}^{\phantom{c}b}\right) {\bf e}_b -
\left(\lambda^{aI}-\Lambda^{abJ}\omega_{bJ}^{\phantom{bJ}I}\right){\bf n}_I\,.
\end{equation}
The Euler-Lagrange equation for ${\bf n}$ is
\begin{equation}
 (\lambda^{aI}+\nabla_b \Lambda^{abI}-2 \Lambda^{bIJ} K_{b\phantom{a}J}^{\phantom{b}a}) {\bf
e}_a-(\nabla_a \Lambda^{aIJ}+\Lambda^{abI}K_{ab}^{\phantom{ab}J}+\lambda^{IJ}){\bf n}_J=0\,.
\end{equation}
Thus from the linear independence of the adapted basis,  one identifies
\begin{equation}
\label{L1}
 \lambda^{aI}=-\nabla_b \Lambda^{abI}+2 \Lambda^{bIJ} K_{b\phantom{a}J}^{\phantom{b}a}\,, \quad
\lambda^{IJ}=-\nabla_a \Lambda^{aIJ}-\Lambda^{abI}K_{ab}^{\phantom{ab}J}\,.
\end{equation}
There remains to determine the three Lagrange multipliers $\lambda^{ab}$, $\Lambda^{abI}$ and
$\Lambda^{aIJ}$. They are determined from the Euler-Lagrange equations for $g_{ab}$, $K_{abI}$ and
$\omega_{a}^{\phantom{a}IJ}$ respectively:
\begin{equation}
\label{L2}
\lambda^{ab}=K_I (K^{abI}-\frac{1}{4} K^I g^{ab})\,; \quad
\Lambda^{abI}=-K^I g^{ab}\,; \quad
\Lambda^{aIJ}=0\,.
\end{equation}
$\Lambda^{aIJ}$ vanishes because the connection
$\omega_{a}^{\phantom{a}IJ}$ does not appear explicitly in the
bending energy. Had the definition of $\omega_{a}^{\phantom{a}IJ}$
been omitted, no error would have been incurred. However, one
would not be so lucky if the energy had depended explicitly on
$\omega_{a}^{\phantom{a}IJ}$.
\vskip1pc \noindent
Substituting the expressions for the Lagrange multipliers given by Eqs. (\ref{L1})
and (\ref{L2}) into Eq.(\ref{falambda}), ${\bf f}^a$ assumes the form
 \begin{equation} \label{eq:fa}
{\bf f}^a = K_I \left(K^{abI} -\frac{1}{2}K^{I}g^{ab}\right) {\bf e}_b
- \tilde{\nabla}^{a}K^{I}{\bf n}_I\,,
\end{equation}
where
$\tilde{\nabla}_a$ is the covariant derivative involving the
normal connection $\omega_{a\phantom{I}J}^{\phantom{a}I}$.

\setcounter{equation}{0}
\renewcommand{\thesection}{Appendix \Alph{section}}
\renewcommand{\thesubsection}{C. \arabic{subsection}}
\renewcommand{\theequation}{C. \arabic{equation}}

\section{Hypersurface adapted Euler-Lagrange derivatives} \label{appELder}

In this appendix,  the details of the derivation of
Eqs.(\ref{Euler1}) and (\ref{Euler2}) are provided. First
dismantle the covariant Laplacian $\tilde\nabla^2K^I$ in terms of
the normal connection, so that
\begin{eqnarray}
\tilde\nabla^2K_I&=&g^{ab} \left(\nabla_a(\tilde\nabla_b K_I)-\omega_{aIJ}\tilde\nabla_bK_J
\right)\nonumber\\
&=& \nabla^2  K_I + \nabla^a \omega_{aIJ} K^J + 2 \omega_{aIJ}\nabla^a
K^J+\omega_{aI}^{\phantom{aI}J}\omega_{aJL}K^L\,.
\end{eqnarray}
The components along ${\bf n}$ and ${\bf l}$ are given respectively by
\begin{subequations}
\begin{eqnarray}
\tilde\nabla^2 K^{(1)}&=&\nabla^2 K_{(1)} + \kappa \nabla \cdot \omega + 2 \omega \cdot \nabla\kappa - |\omega|^2  K_{(1)}\,;
\label{Auno}\\
\tilde\nabla^2\kappa&=& \nabla^2 \kappa - K_{(1)}\nabla \cdot \omega - 2 \omega\cdot \nabla  K_{(1)} - |\omega |^2 \kappa\,.
\label{Ados}
\end{eqnarray}
\end{subequations}
where $\nabla \cdot \omega = \nabla^a \omega_{a12}$ is the
divergence of the normal connection and
$|\omega|^2=\omega^a_{\phantom{a}12}\omega_a^{\phantom{a}12}$ is
its squared modulus. For an embedding in a hypersphere, the normal
connection vanishes, $\omega_{a12}=0$, so that the normal
covariant derivatives are replaced by covariant derivatives,
$\tilde \nabla^2 \kappa= \nabla^2\kappa$; in addition, $K_{(1)}$
is constant, so that $\tilde\nabla^2 K_{(1)}=0$.
\vskip1pc\noindent More generally, Eq. (\ref{omega12}) is used to
express the divergence and squared modulus of the normal
connection in terms of hypersurface curvatures. The second term is
given by
\begin{eqnarray} \label{sndtrmAdos}
\nabla\cdot \omega :&=& \nabla^a\omega_{a12}=-\nabla^a( {\cal E}_a^Al^BK_{AB})\nonumber\\
&=& \kappa l^A l^B K_{AB} - \kappa^{ab} {\cal E}_a^A {\cal E}_b^B K_{AB} - g^{ac} {\cal E}_a^A {\cal
E}_c^C l^B \nabla_C K_{AB}
\nonumber\\
&=&\kappa l^A l^B K_{AB} - \kappa^{ab}K_{ab\,(1)}-{\cal H}^{AB} l^C\nabla_C K_{AB} \,.
\end{eqnarray}
Here ${\cal H}^{AB}=g^{ab}{\cal E}_a^A{\cal E}_b^B$ represents the
projector onto $\Gamma_{N,N+1}$. On the second line, the relations
$\nabla^a l^B=\kappa^{ab}{\cal E}_b^B$ and $D^a{\cal
E}_a^A=-\kappa l^A$ are used, which follow from the structure
equations associated with the embedding of the surface
$\Gamma_{N,N+1}$ in $\Sigma_{N+1}$, (Eq. (\ref{G23}) and its
contraction). In addition, the Codazzi-Mainardi equation for
$\Sigma_{N+1}$, $\nabla_AK_{BC}=\nabla_BK_{AC}$ has been used in
the third line.
\vskip1pc \noindent
The squared modulus is given by
\begin{eqnarray}
|\omega |^2 :&=& g^{ab}\omega_{a12}\omega_{b12} = g^{ab}{\cal E}_a^A{\cal
E}_b^B l^C l^D K_{AC} K_{BD}\nonumber\\
&=&l^Al^B \left(K_{AC} K_B^{\phantom{B}C} -K_{AC}K_{BD} l^C l^D\right)\nonumber\\
&=& l^A l^B \left(K_{(1)} K_{AB} - R_{AB}\right)\,,\label{Atres}
\end{eqnarray}
where the completeness of the basis $\{{\cal E}^A_a,l^A\}$ and the
contracted Gauss-Codazzi equation have been used. Here $R_{AB}$ is
the Ricci tensor constructed with the metric $G_{AB}$.
\vskip1pc\noindent
One also requires the decomposition of  $K_{ab}^{\phantom{ab}I} f^{ab}$ where $f^{ab}={\bf f}^a \cdot
{\bf e}^b=K_I (K^{abI}-1/2 g^{ab}K^I)$, which are given by
\begin{subequations}
\begin{eqnarray}
K_{ab}^{\phantom{ab}(1)}f^{ab} &=& K^{(1)} \left(K_{ab\,(1)} K^{ab\,(1)}-\frac{1}{2}
K_{(1)}^2 -\mu \right) + \kappa\left( \kappa_{ab} K^{ab\,(1)} - \frac{1}{2}\kappa K^{(1)} \right)
\,; \label{Kabfab1}\\
\kappa_{ab}f^{ab} &=& K^{(1)}\left(\kappa_{ab} K^{ab\,(1)} - \frac{1}{2}\kappa K^{(1)}\right) +
\kappa\left(\kappa_{ab}\kappa^{ab}-\frac{1}{2}\kappa^2 -\mu\right)\,. \label{kappaabfab}
\end{eqnarray}
\end{subequations}
For a surface embedded in a hypersphere, with
$K_{ab}^{\phantom{ab}(1)}= g_{ab}$, $\kappa \left(\kappa_{ab}
K^{(1)\, ab}-\kappa K^{(1)}/2\right) = -(N-2)\kappa^2 /2 $ and
$K^{(1)}\left(K_{ab\,(1)} K^{ab\,(1)} - K_{(1)}^2/2\right) =
-N^2(N-2)/2$. \vskip1pc\noindent For confined curves, with
$K^{(1)}=\kappa_n$ and $\kappa=\kappa_g$,  Eq, (\ref{Kabfab1})
reduces to $\kappa_n [(\kappa_g^2+\kappa_n^2)/2 -\mu]$, whereas
Eq.(\ref{kappaabfab}) reduces to $\kappa_g
[(\kappa_g^2+\kappa_n^2)/2-\mu]$.

\setcounter{equation}{0}
\renewcommand{\thesection}{Appendix \Alph{section}}
\renewcommand{\thesubsection}{D. \arabic{subsection}}
\renewcommand{\theequation}{D. \arabic{equation}}

\section{Variational principle adapted to a hypersphere} \label{apphypbend}

It is instructive to decompose the bending energy into parts adapted to its confined environment.
Using Eqs.(\ref{K1}) and (\ref{G23}) one finds
\begin{equation} \label{Hamk3}
H = \frac{1}{2}\, \int dA_N \, K^{(I)} K_{(I)} =\frac{1}{2}\, \int
dA_N\, [ \kappa^2 +  (K - l^A l^B K_{AB})^2 ]\,.
\end{equation}
The first term is the bending energy of the
surface embedded into the three-dimensional curved geometry described by the metric tensor $G_{AB}$,
\begin{equation} \label{Hamk4}
H_0 = \frac{1}{2}\, \int dA_N\, \kappa^2 \,.
\end{equation}
The second is the bending energy inherited from the hypersurface.
This is the generalization to higher dimensions of the Pythagorean
decomposition of the Frenet curvature $k$ of a space curve into
geodesic and normal parts, $k^2 = \kappa_g^2 + \kappa_n^2$
familiar in  the theory of surfaces \cite{DoCarmo}.
\vskip1pc\noindent The Euler Lagrange equation for the surface
$s\to U(s)$  described by the bending energy (\ref{Hamk4}) is
given by
\begin{equation}
\label{eulerkappa} -\, \nabla^2 \kappa  + \frac{1}{2} (\kappa^2- 2
\kappa_{ab} \kappa ^{ab}) \kappa - R_{AB} l^A l^B \kappa + \mu
\kappa =0\,.
\end{equation}
This equation was derived in the 90s in the context of relativistic extended
objects \cite{CapGuv}.
\vskip1pc \noindent
For an $N+1$-sphere, the Ricci tensor is
proportional to the metric tensor, $R_{AB}= N  G_{AB}$ so that
$R_{AB} l^A l^B = N$.\footnote{ The Gauss-Codazzi equations
describing the embedding of the hyperface in the Euclidean
background, $R_{AB}= K K_{AB}- K_{AC} K^C{}_B$, together with the
identity, $K_{AB}= G_{AB}$, are useful for remembering the
relationship between $R_{AB}$ and $G_{AB}$. Of course this
relationship is completely intrinsic.} In this case $K_{AB}=
G_{AB}$, and $K=N+1$ so that Eq. (\ref{Hamk3}) then reduces to
\begin{equation} \label{Hamkappa}
H = \frac{1}{2}\, \int dA_N\, [ \kappa^2 +  N^2]\,.
\end{equation}
The contribution to the energy density due to the normal curvature is constant. If the
Euler-Lagrange derivative corresponding to this contribution ($N^2
\kappa/2$) is added on the left hand side of Eq.
(\ref{eulerkappa}),  the shape equation Eq. (\ref{elsph}) is
reproduced.

\setcounter{equation}{0}
\renewcommand{\thesection}{Appendix \Alph{section}}
\renewcommand{\thesubsection}{E. \arabic{subsection}}
\renewcommand{\theequation}{E. \arabic{equation}}

\section{Rotational invariance and confinement by a hypersphere} \label{approtation}

If the confining surface is a hypersphere, the Euler Lagrange
equations will respect rotational invariance in
$\mathbb{R}^{4}$. Thus, whereas stress is not conserved, torques
will be, corresponding as they do to the Noether current
associated with the rotational symmetry of the energy. \vskip1pc
\noindent Consider a small rotation about the origin
$\delta X^i=\epsilon^{ijkl}\Omega^{jk}X^l$, characterized by
the antisymmetric rotation matrix $\Omega^{ij}$, where
$\epsilon^{ijkl}$ is the Levi-Civita symbol and $i,j,k,l=
1,\dots,4$ index tensors in Euclidean space $\mathbb{R}^{4}$.
Each of the two normal vectors ${\bf n}^I, I=1,2$ rotates
accordingly:  $\delta {n}^{Ii}=\epsilon^{ijkl}\Omega^{jk}n^{Il}$.
It can be shown that, in equilibrium, on any surface patch
\begin{equation}
\delta H_c = - \Omega^{ij} \, \int dA_2 \, \nabla_a m^{ij\, a}\,,
\end{equation}
where
\begin{equation}
m^{aij} =\epsilon^{ijkl}\left( f^{ak}X^l-K_Ig^{ab}e^k_bn^{Il}\right)\,.
\end{equation}
Here $f^{ak} = {\bf f}^a \cdot \hat{\bf x}^k$, $e_{a}^k = {\bf e}_a
\cdot \hat{\bf x}^k$, etc.
\vskip1pc \noindent
Thus $m^{ij\, a}$ is conserved if $H_c$ is rotationally invariant.  This is confirmed by the following calculation:
\begin{equation}
\nabla_a m^{aij} =\epsilon^{ijkl}\left( \nabla_a f^{ak}X^l +  f^{ak} e^l_a -K_I  K^{I ab} e^k_a
e^l_b - \tilde{\nabla}_a K_I g^{ab} e^k_b n^{Il} - K_I K_J n^{Jk} n^{Il} \right)\,.
\end{equation}
In general, all but the first term either vanish identically or
cancel among themselves.  For the first term one has
$\epsilon^{ijkl}\nabla_a f^{ak}X^l = \epsilon^{ijkl}
\varepsilon_{\bf n} n^k X^l$. However, if the surface is confined to
a unit hypersphere centered on the origin, so that ${\bf X} = {\bf
n}$, this term also vanishes.
\vskip1pc \noindent
It is not obvious how to exploit this invariance in a manner analogous to that in the treatment of a
polymer confined within a sphere. In any case a better strategy is
to exploit the conformal invariance of the two-dimensional bending
energy, as sketched in section \ref{Discussion}.

\end{appendix}

\end{document}